\documentclass[pdflatex,sn-apa]{sn-jnl}

\usepackage{graphicx}%
\usepackage{multirow}%
\usepackage{amsmath,amssymb,amsfonts}%
\usepackage{amsthm}%
\usepackage{mathrsfs}%
\usepackage{xcolor}%
\usepackage{textcomp}%
\usepackage{manyfoot}%
\usepackage{booktabs}%
\usepackage{algorithm}%
\usepackage{algorithmicx}%
\usepackage{algpseudocode}%
\usepackage{listings}%
\usepackage{comment}%

\def\b#1{\mbox{\boldmath $#1$}}    

\theoremstyle{thmstyleone}%
\theoremstyle{thmstyletwo}%

\theoremstyle{thmstylethree}%

\begin{document}

\title[Article Title]{Variable selection via knockoffs for clustered data}


\author*[1]{\fnm{Silvia} \sur{Bacci}}\email{silvia.bacci@unifi.it}

\author[1]{\fnm{Leonardo} \sur{Grilli}}\email{leonardo.grilli@unifi.it}

\author[1]{\fnm{Carla} \sur{Rampichini}}\email{carla.rampichini@unifi.it}
\affil*[1]{\orgdiv{Department of Statistics, Computer Science, Applications `G.Parenti'}, \orgname{University of Florence}, \orgaddress{\street{Viale Morgagni,59}, \city{Firenze}, \postcode{50134}, \state{}, \country{Italy}}}


\abstract{
We extend the knockoffs method for selecting predictors in the case of clustered data (cross-section or repeated measures). In the setting of clustered data, variable selection is complex because some predictors are measured at the observation level (level 1), whereas others are measured at the cluster level (level 2), so their values are constant within clusters. The solution we propose is to conduct variable selection separately at the two levels. To this end, we suggest a two-step approach: (i) decompose each level 1 predictor into level 2 and level 1 components by replacing it with the cluster mean and the deviation from the cluster mean; (ii) perform variable selection separately at the two levels, where the level 1 data matrix includes the deviations from the cluster means and the level 2 data matrix includes the cluster means of level 1 predictors and the level 2 predictors.
To evaluate the performance of the proposed approach, we conduct a simulation study comparing the sequential knockoff, the derandomized knockoff, and the Lasso. The study shows satisfactory results in terms of false discovery rate and power. All methods fail when applied to the complete data matrix, including both level 1 and level 2 predictors. In contrast, all methods perform better when applied to the level 1 and level 2 data matrices separately. Moreover, the sequential knockoffs method performs substantially better than the Lasso and the derandomized knockoffs. Our proposal to implement the knockoffs method in a clustered data framework is feasible, flexible, and effective.}




\maketitle

\section{Introduction}\label{sec1}

The selection of relevant predictors affecting a response is a fundamental issue in assessing a statistical model. It is particularly challenging in high-dimensional contexts, when numerous predictors are available. Indeed, different selection strategies may yield different results, with the risk of including variables with null effects in the model or, conversely, excluding variables with non-null effects.
The knockoffs approach \citep{barber:candes:2015} has the advantage of explicitly controlling either the Per-Family Error Rate (PFER), which is the expected number of false discoveries, or the False Discovery Rate (FDR), which is the expected proportion of false discoveries.

The available knockoff procedures are not explicitly designed for complex data structures, such as clustered (or hierarchical) data, a setting often encountered in applications, such as panel data (waves nested within subjects), repeated measures (occasions nested within subjects), or when data are collected from multiple groups (e.g., patients within hospitals). In such a setting,  linear mixed models \citep{snijders:2011} are a valuable extension of simple linear models that allow for properly taking into account the within-group correlation of observations arising from hierarchical data. In this contribution, we focus on the linear random intercept model for a continuous response. This model includes predictors measured with fixed effects at both hierarchical levels and a random intercept to account for the unobserved heterogeneity among clusters.

The variable selection in the linear random intercept model is more complex than the simple linear model, because some predictors are measured at the observation level (level 1), whereas others are measured at the cluster level (level 2), with their values being constant within clusters by definition.
Unfortunately, the available procedures for variable selection via the knockoff filter are not suitable for clustered data, because the algorithms for generating knockoffs cannot account for the restriction of constant values of level 2 predictors within the clusters.

In this contribution, we propose a strategy to extend the knockoff filter to clustered data by selecting the predictors separately at the two hierarchical levels.

The rest of the paper is organized as follows. Section \ref{sec:methodback} briefly reviews the basic principles of the knockoff filter, while Section \ref{sec:methodproposal} illustrates the strategy for extending the procedure to the case of clustered data.
Section \ref{sec:sim} presents a simulation study to evaluate the performance of the proposed strategy.  Section \ref{sec:final} concludes with some remarks.

\section{Variable selection via knockoffs: methodological background}
\label{sec:methodback}
The knockoff filter is a recently proposed method for selecting the set of {\em non-null} variables, that is, the predictors that influence the outcome in the population model \citep{barber:candes:2015,candes:2018}.
This approach, which has been developed for high-dimensional settings, has the advantage of explicitly controlling either the PFER or the FDR.

In a regression model for the outcome $Y$ on a set of predictors $\mathbf X$, the knockoffs are copies $\mathbf X^*$ of the predictors with two properties: (i) given the observed variables, the copies are unrelated to the outcome, and (ii) $\mathbf X^*$ and $\mathbf X$ have the same distribution.  More precisely, given $K$ observed predictors $\mathbf X=[X_1, \ldots, X_K]$, the knockoffs ${\mathbf X}^*=[{X}_1^*, \ldots, {X}_K^*]$ are randomly generated variables independent of $Y$ conditionally on $\mathbf X$ and with the following swapping property, holding for any subset $S$ of predictors:

\begin{equation}
 \label{eq:swap}
 [\mathbf X, {\mathbf X}^*]_{swap(S)} \overset{d}{=} [\mathbf X, {\mathbf X}^*]
 \end{equation}
where $S \subseteq \{1, \ldots, K\}$ and $\overset{d}{=}$ denotes the equality in distribution. The vector $[\mathbf X, {\mathbf X}^*]_{swap(S)}$ is obtained by swapping the entries $X_k$ and ${X}_k^*$ for any $k \in S \subseteq \{1, \ldots, K\}$, that is, the distribution of the augmented vector $[\mathbf X, {\mathbf X}^*]$ does not change if the original variables are swapped with the corresponding knockoff copies.

After the generation step, the variable selection step is performed by regressing the outcome $Y$ on the augmented vector of predictors $[\mathbf X, {\mathbf X}^*]$, using a regularization procedure such as the Lasso \citep{tibshirani:1996}. Then, for each element $X_k$ in $\mathbf X$, a statistic is computed, for example, the difference between the absolute value of the estimated coefficient for the original variable and the corresponding estimate for its knockoff copy \citep{barber:candes:2015}.

In the original knockoffs approach, the set of selected variables may depend on the specific values randomly generated for the knockoffs ${\mathbf X}^*$. The derandomized knockoffs approach \citep{ren:2023} was proposed to overcome the non-uniqueness of the knockoff solution by repeating the generation-selection procedure a given number of times (e.g., $31$). Then, variables chosen in at least a given proportion of cases are selected (usually, $0.50$).

The knockoff method was further extended to deal with categorical predictors using a sequential procedure \citep{kormaksson:2021}. According to this approach, knockoffs are generated differently depending on the variable type: sampling from a normal distribution for continuous variables and from a multinomial distribution for categorical ones. The parameters of the distributions are estimated sequentially by regressing each predictor on the remaining variables and the previously generated knockoffs, using penalized (linear or multinomial logistic) models, as appropriate. Once the knockoffs are created, the regression coefficients of the original and knockoff variables are compared through suitable statistics, following the same approach as \cite{candes:2018}. As in \cite{ren:2023}, the generation-selection procedure repeats multiple times. The sequential knockoffs procedure has been recently optimized computationally by \cite{zimmermann:2024} through the development of the sparse sequential knockoff algorithm, which introduces a preliminary phase based on graphical Lasso aimed at pre-selecting predictors.

\section{Implementing the knockoff filter for clustered data}
\label{sec:methodproposal}

In the setting of clustered data, variable selection becomes more complex because some predictors are measured at level 1, while others are measured at level 2. Level 2 predictors are constant within clusters by definition, a feature not accounted for by current methods to generate knockoffs. To overcome this limitation, we propose separating the generation of knockoffs and subsequent variable selection at the two levels.

To this end, in a linear model, we need the predictors at level 1 to be uncorrelated to those at level 2, since in this way the least squares regression coefficients of the model with all predictors are the same as the coefficients obtained from the level 1 model and the level 2 model separately \citep[Section 3.3]{greene:2003}.

Uncorrelated predictors among level 1 and level 2  can be obtained by centering level 1 predictors around their cluster means.
Formally, we denote with $j=1, \ldots, J$ the level 2 units (clusters) and with $i=1, \ldots, n_j$ the level 
1 units. Then, we denote $X_{ij}$ as a level 1 predictor and $Z_{j}$ as a level 2 predictor.
The cluster mean of a level 1 predictor is $\bar{X}_{j} = \sum_{i = 1}^{n_j}X_{ij}/n_j$, so the centered version is $\tilde{X}_{ij} = X_{ij} - \bar{X}_{j}$. By construction, the centered level 1 predictor $\tilde{X}_{ij}$ is uncorrelated with any variable that is constant within the clusters, such as a level 2 predictor $Z_j$ or the cluster mean of a level 1 predictor $\bar{X}_j$. Indeed, since  $\sum_{i = 1}^{n_j}\tilde{X}_{ij} =0$, then $Cov(\tilde{X}_{ij}, Z_{j})=Cov(\tilde{X}_{ij}, \bar{X}_{j})=0$.

It is worth noting that the regression coefficient of a level 1 predictor can be decomposed into a within coefficient associated with the centered component $\tilde{X}_{j}$ and a between coefficient associated with the cluster mean $\bar{X}_{j}$. If the within and between coefficients differ in the population, a model with only the original predictor $X_{ij}$ is misspecified, as it provides an estimated regression coefficient that is a weighted average of the within and between coefficients \citep{snijders:2011}. Therefore, the proposed strategy of centering level 1 predictors is good practice regardless of the aim of generating knockoffs.

In our proposal, the generation of the knockoffs and the subsequent variable selection are carried out separately for level 1 and level 2 predictors.
Specifically, we select level 1 predictors by regressing the response $Y_{ij}$ on the level 1 centered predictors $\tilde{X}_{ij}$. On the other hand, for the selection of level 2 predictors, we aggregate the data by cluster and regress the average response $\bar{Y}_{j} = \sum_{i = 1}^{n_j}Y_{ij}/n_j$ on $Z_{j}$ and possibly the cluster means of level 1 predictors $\bar{X}_{j}$, by a weighted regression with $J$ observations and cluster sizes $n_j$ as weights. In this way, we obtain the same regression coefficients as if we had regressed the response $Y_{ij}$ on $Z_{j}$ and $\bar{X}_{j}$ in the full dataset, which contains $\sum_j n_j$ observations.

The proposed approach is particularly appealing to practitioners because it can be implemented using the available procedures for variable selection via knockoffs, with the only warning to insert weights $n_j$ for the selection of level 2 predictors, except when the clusters have the same size, i.e., $n_j=n$.

Note that the false discovery measures (PFER or FDR) are controlled separately at each level. This is an advantage, especially if the number of predictors differs significantly at the two levels and the aim is to control the PFER.

\section{Monte Carlo simulation study}
\label{sec:sim}
This section presents the results of a Monte Carlo study evaluating the performance of the proposed method for applying the knockoff filter to clustered data. We first describe the simulation setup and then discuss the main findings.

\subsection{Simulation setup}
In the Monte Carlo study, we simulate 100 samples with a repeated measures structure.
Each simulated sample contains $1500$ observations and $60$ predictors. The observations correspond to $J = 300$ individuals (i.e., clusters or level 2 units). Each individual has  $n_j = 5$ repeated measurements (i.e., level 1 observations). The predictors comprise $K = 20$ time-varying variables $X_{kij}$ ($k = 1, \ldots, K$; $i = 1, \ldots, n_j$; $j = 1, \ldots, J$), the corresponding 20 cluster means $\bar{X}_{k.j} = \sum_{i = 1}^{n_j} X_{kij}/n_j$, and $H = 20$ time-constant variables $Z_{hj}$ ($h = 1, \ldots, H$; $j = 1, \ldots, J$). In total, there are 20 level 1 variables ($X_{kij}$) and 40 level 2 variables ($\bar{X}_{k.j}$ and $Z_{hj}$).

Level 2 variables are generated from a multivariate Normal distribution with mean $\b 0$ and correlation structure specified similarly to \cite{xie:2023}. Specifically, the 40 level 2 predictors are divided into four blocks, each containing five variables $\bar{X}_{k.j}$ and five variables $Z_{hj}$ (i.e., block 1: $h,k = 1,\ldots,5$; block 2: $h,k = 6,\ldots,10$; block 3: $h,k = 11,\ldots,15$; block 4: $h,k = 16,\ldots,20$). Within-block correlations are 0.60 for the first three blocks and 0.30 for the fourth block, whereas between-block correlations are either 0.30 or 0.15. Table \ref{tab:corrmatr_lv2} illustrates the correlation structure for a simplified case with two variables per block.

\begin{table}[!ht]
\caption{Upper-half correlation matrix of level 2 variables (example with two variables per block).}\label{tab:corrmatr_lv2}%
\begin{tabular}{ccccc}
\toprule
 & Block 1 & Block 2 & Block 3 & Block 4 \\
 & $\bar{X}_{1.j} \;\; Z_{1j}$ & $\bar{X}_{6.j} \;\; Z_{6j}$ & $\bar{X}_{11.j} \;\; Z_{11j}$ & $\bar{X}_{16.j} \;\; Z_{16j}$ \\
\midrule \vspace{2mm}
Block 1 $\begin{array}{c}
\bar{X}_{1.j}  \\
Z_{1j} \\
\end{array} $ &
$\left(\begin{array}{cc}
1.00 & 0.60  \\
0.60 & 1.00  \\
\end{array}\right) $ &
$\left(\begin{array}{cc}
0.30 & 0.15  \\
0.15 & 0.30  \\
\end{array}\right) $ &
$\left(\begin{array}{cc}
0.15 & 0.15  \\
0.15 & 0.15  \\
\end{array}\right) $ &
$\left(\begin{array}{cc}
0.30 & 0.15  \\
0.15 & 0.30 \\
\end{array}\right) $ \\\vspace{2mm}
Block 2 $\begin{array}{c}
\bar{X}_{6.j}  \\
Z_{6j} \\
\end{array} $ &  & $\left(\begin{array}{cc}
1.00 & 0.60  \\
0.60 & 1.00  \\
\end{array}\right) $ & $\left(\begin{array}{cc}
0.15 & 0.15  \\
0.15 & 0.15  \\
\end{array}\right) $ & $\left(\begin{array}{cc}
0.30 & 0.15  \\
0.15 & 0.30 \\
\end{array}\right) $ \\\vspace{2mm}
Block 3 $\begin{array}{c}
\bar{X}_{11.j}  \\
Z_{11j} \\
\end{array} $ &  &  & $\left(\begin{array}{cc}
1.00 & 0.60  \\
0.60 & 1.00  \\
\end{array}\right) $ &  $\left(\begin{array}{cc}
0.15 & 0.15  \\
0.15 & 0.15  \\
\end{array}\right) $\\\vspace{2mm}
Block 4 $\begin{array}{c}
\bar{X}_{16.j}  \\
Z_{16j} \\
\end{array} $ &  &  &  & $\left(\begin{array}{cc}
1.00 & 0.30  \\
0.30 & 1.00  \\
\end{array}\right) $ \\
\botrule
\end{tabular}
\end{table}
Next, two time-constant variables $Z_{hj}$ for each block are dichotomized. Thus, among the 40 level 2 predictors,  32 are continuous (normal)  and 8 are binary ($h = 1, 2, 6, 7, 11, 12, 16, 17$).

Level 1 variables $X_{kij}$ are generated from their cluster means $\bar{X}_{k.j}$ using a linear regression model with a linear time effect $\gamma$ and an error term $v_{ij}$:
\begin{equation}
X_{kij} = \bar{X}_{k.j} + \gamma\cdot time_i + v_{ij},
\label{eq:genX}
\end{equation}
where  $time_i=\{-2,-1, 0,1,2\}$ and  $v_{ij}$ has a multivariate Normal distribution with mean $\b 0$ and correlation structure specified in Table \ref{tab:corrmatr_lv1}.

\begin{table}[!ht]
\caption{Upper-half correlation matrix of the error component of level 1 variables.}\label{tab:corrmatr_lv1}%
\begin{tabular}{ccccc}
\toprule
 & $v_{1ij} \ldots v_{6ij}$ & $v_{7ij} \ldots v_{16ij}$ & $v_{17ij} \ldots v_{20ij}$ \\
\midrule \vspace{2mm}
$\begin{array}{c}
v_{1ij}  \\
\vdots \\
v_{6ij} \\
\end{array} $ &
$\left(\begin{array}{ccc}
1.00 & \ldots & 0.60  \\
\vdots & \ddots  & \vdots \\
0.60 & \ldots & 1.00  \\
\end{array}\right) $ &
$\left(\begin{array}{ccc}
0.15 & \ldots& 0.15  \\
\vdots & \ddots&    \vdots \\
0.15 & \ldots& 0.15  \\
\end{array}\right) $ &
$\left(\begin{array}{ccc}
0.60 & \ldots & 0.60  \\
\vdots & \ddots &    \vdots \\
0.60 & \ldots & 0.60  \\
\end{array}\right) $ \\ \vspace{2mm}
$\begin{array}{c}
v_{7ij}  \\
\vdots \\
v_{16ij} \\
\end{array} $ &
 & $\left(\begin{array}{ccc}
1.00 & \ldots & 0.30  \\
\vdots & \ddots &    \vdots \\
0.30 & \ldots & 1.00  \\
\end{array}\right) $ & $\left(\begin{array}{ccc}
0.15 & \ldots & 0.15  \\
\vdots & \ddots &    \vdots \\
0.15 & \ldots & 0.15 \\
\end{array}\right) $ \\\vspace{2mm}
$\begin{array}{c}
v_{17ij}  \\
\vdots \\
v_{20ij}\\
\end{array} $ &
 &   &
$\left(\begin{array}{ccc}
1.00 & \ldots & 0.60  \\
\vdots & \ddots &    \vdots \\
0.60 & \ldots & 1.00  \\
\end{array}\right) $\\
\botrule
\end{tabular}
\end{table}

The response variable $Y_{ij}$ is generated by a random intercept linear regression model as
\begin{equation}
Y_{ij} = \sum_{k = 1}^K \beta_{W,k} \tilde{X}_{kij} + \sum_{k = 1}^K \beta_{B,k} \bar{X}_{k.j} + \sum_{h = 1}^H \delta_{h} Z_{hj} + u_j + e_{ij},
\label{eq:genY}
\end{equation}
with $\tilde{X}_{kij} = X_{kij} - \bar{X}_{k.j}$.
We assume independent errors $u_j \sim N(0, 4)$ and $e_{ij} \sim N(0, 1)$,  corresponding to an intraclass correlation coefficient $ICC=4/(4+1)=0.80$. Regression coefficients $\beta_{W,k}$,  $\beta_{B,k}$, and $\delta_{h}$ are set to 0 for null variables and to 0.5 or 1.0 for non-null variables. Specifically, we assume a total of 12 non-null variables: 4 at level 1 (i.e., $\beta_{W,1} = \beta_{W,7} = 1.0$ and $\beta_{W,2} = \beta_{W,8} = 0.5$) and 8 at level 2, of which 6 continuous (i.e., $\beta_{B,1} = \beta_{B,13} = \delta_{13} = 1.0$;  $\beta_{B,7} = \beta_{B,19} = \delta_{19} = 0.5$) and 2 binary  (i.e., $\delta_{1} = 1.0$ and $\delta_{7} = 0.5$).

In the simulation study, we apply the variable selection procedure described in Section \ref{sec:methodproposal} using two methods: (1) the derandomized knockoff filter \citep{ren:2023} with 31 runs and selection threshold 0.5, (2) the sparse sequential knockoff filter introduced by \cite{kormaksson:2021} and later optimized by  \cite{zimmermann:2024}. Since the derandomized knockoff filter allows to control only the PFER, to ensure comparability between the two methods, we set the nominal value of the PFER, using either PFER=$1$ or PFER$=2$.

We also compare the two knockoff methods with a standard variable selection procedure based on the elastic net, considering two values of mixing parameter, $\alpha = 1$ (standard Lasso) and $\alpha = 0.5$. We use two criteria to select the optimal regularization parameter $\lambda$: (\emph{i}) the value of $\lambda$ that minimizes the mean cross-validated error, and (\emph{ii}) a more parsimonious choice corresponding to the largest $\lambda$ for which the mean cross-validated error is within one standard error of the minimum.

All analyses are conducted in {\tt R}, using the following packages: \texttt{derandomKnock} \citep{derandomKnock} for the derandomized knockoff,
\texttt{knockofftools} \citep{zimmermann:2024} for the sparse sequential knockoff, and \texttt{glmnet} \citep{glmnet} for the elastic net.

\subsection{Simulation results}
\label{sec:sim_2}
Following our proposal (Section \ref{sec:methodproposal}), the three variable selection approaches (elastic net, derandomized knockoffs, and sparse sequential knockoffs) are applied on level 1 and level 2 datasets separately.
The level 1 dateset has  $1500$  rows ($n_j=5$ repeated measures by $J=300$ individuals) and $21$ columns, i.e., the $K=20$ time-varying centered variables $\tilde{X}_{kij}$ plus the response variable $Y_{ij}$; (ii) the level 2 dataset has $J = 300$ rows and $41$ columns, i.e., the $K=20$ cluster mean variables $\bar{X}_{k.j}$, the $H=20$ time-constant variables $Z_{hj}$,  and the cluster mean response variable $\bar{Y}_{.j} = \sum_{i = 1}^{n_j} Y_{ij}/n_j$.

Table \ref{tab:sim1} presents the results of the simulation study, expressed as Monte Carlo (MC) means of empirical PFER, FDR, and True Positive Rate (TPR), for each of the two hierarchical levels.
Specifically, the MC mean of PFER ($\widehat{PFER}$) is the average number of null variables selected by the procedure; the MC mean of FDR ($\widehat{FDR}$) is the average proportion of  null variables selected, out of the total number of selected variables;  the MC mean of the TPR ($\widehat{TPR}$) is the average proportion of non-null variables selected by the method.

\begin{table}[!ht]
\caption{Variable selection with elastic net, derandomized knockoff, and sparse sequential knockoff, by hierarchical level: Monte Carlo means of  $\widehat{PFER}$, $\widehat{FDR}$, and $\widehat{TPR}$ (nominal $PFER = 1, 2$)}\label{tab:sim1}
\begin{tabular*}{\textwidth}{@{\extracolsep\fill}lcccccc}
\toprule%
Method	&	 PFER	&	$\widehat{PFER}$	&	$\widehat{FDR}$	&	$\widehat{TPR}$	\\
\midrule
{\em Selection at level 1}	&		&		&		&		\\
{(4 non-null predictors out of 20; 1500 obs)}	&		&		&		&		\\
 elastic net 	&		&		&		&		\\
$\;\;$ $\alpha = 1$; min $\lambda$	&	--	&	5.76	&	0.55	&	1.00	\\
$\;\;$ $\alpha = 1$; min $\lambda$ + 1 s.e.	&	--	&	0.77	&	0.13	&	0.99	\\
$\;\;$ $\alpha = 0.5$; min $\lambda$	&	--	&	6.62	&	0.59	&	1.00	\\
$\;\;$ $\alpha = 0.5$; min $\lambda$ + 1 s.e.	&	--	&	1.87	&	0.28	&	1.00	\\
 derandomized knockoff 	&	1	&	1.16	&	0.19	&	1.00	\\
 sequential knockoff 	&	1	&	0.55	&	0.10	&	0.99	\\
 derandomized knockoff 	&	2	&	2.41	&	0.33	&	1.00	\\
 sequential knockoff 	&	2	&	1.60	&	0.25	&	1.00	\\
\midrule
{\em Selection at level 2}	&		&		&		&		\\
{(8 non-null predictors out of 40; 300 obs)}	&		&		&		&		\\
elastic net 	&		&		&		&		\\
$\;\;$ $\alpha = 1$; min $\lambda$	&	--	&	11.22	&	0.57	&	0.96	\\
$\;\;$ $\alpha = 1$; min $\lambda$ + 1 s.e.	&	--	&	5.50	&	0.41	&	0.91	\\
$\;\;$ $\alpha = 0.5$; min $\lambda$	&	--	&	12.62	&	0.60	&	0.97	\\
$\;\;$ $\alpha = 0.5$; min $\lambda$ + 1 s.e.	&	--	&	8.18	&	0.51	&	0.93	\\
 derandomized knockoff 	&	1	&	2.08	&	0.21	&	0.87	\\
 sequential knockoff 	&	1	&	0.61	&	0.08	&	0.76	\\
 derandomized knockoff 	&	2	&	3.67	&	0.32	&	0.91	\\
 sequential knockoff 	&	2	&	1.46	&	0.17	&	0.82	\\
\botrule
\end{tabular*}
\end{table}

Table \ref{tab:sim1} shows that the sparse sequential knockoff approach performs well at both levels. In each case, the PFER is properly controlled, with the empirical values below the nominal counterparts. Specifically, at level 1, we obtain $\widehat{PFER} = 0.55$ and $1.60$ for nominal PFER values of 1 and 2, respectively; similar results ($0.61$ and $1.46$) are observed at level 2. In contrast, the derandomized knockoff approach performs less satisfactorily. At level~1, the empirical PFER slightly exceeds the nominal value ($1.16$ vs.\ 1 and $2.41$ vs.\ 2), while at level~2 the gap widens markedly ($2.08$ vs.\ 1 and $3.67$ vs.\ 2), indicating clear difficulty in controlling the PFER.

The elastic net is not designed to control the PFER. Notwithstanding, at level 1, the MC mean values of  PFER for the elastic net with the largest $\lambda$ for which the mean
cross-validated error is within one standard error of the minimum yield results comparable to those of the sparse sequential knockoff. However, at level 2, where the number of predictors is double, the performance is markedly worse.

The comparison among the three methods is similar when looking at the $\widehat{FDR}$.

All three methods show a $\widehat{TPR}$ at level 1 close to 1, meaning that nearly all non-null predictors are correctly selected. At level 2, the elastic net maintains a high TPR at the cost of an extremely large FDR.
When comparing the sparse sequential knockoff with the derandomized knockoffs, a clear trade-off is observed between FDR and TPR. In contrast to the derandomized knockoffs, the sparse sequential knockoff maintains the PFER below the nominal value, albeit at the expense of a lower TPR.
The worst performance of the variable selection methods at  level 2 is a consequence of the different data structure: indeed, at level 1 there are $1500$ observations and $20$ predictors, while at level 2 there are $300$ observations (clusters) and $40$ predictors. A further difference is that at the cluster level, $8$ out of $40$ predictors are binary. However, we also tried a configuration with all continuous predictors (not reported in the paper), obtaining similar results.

To fully appreciate our proposal based on splitting the selection at the two levels, Table \ref{tab:sim2} presents in the top panel the results obtained by applying the variable selection procedures without splitting the selection (referred to as ``overall selection''), whereas the bottom panel of the table presents the results obtained by combining the level 1 and level 2 results of Table \ref{tab:sim1}.

\begin{table}[!ht]
\caption{Variable selection with elastic net, derandomized knockoff, and sparse sequential knockoff: Monte Carlo means of  $\widehat{PFER}$, $\widehat{FDR}$, and $\widehat{TPR}$, for overall selection  ($PFER =  2$)  and for combined results of Table \ref{tab:sim1} ($PFER = 1$ at both levels). Non-null predictors: 12 out of 60; observations: 1500. }\label{tab:sim2}
\begin{tabular*}{\textwidth}{@{\extracolsep\fill}lcccccc}
\toprule%
Method	&	 PFER	&	$\widehat{PFER}$	&	$\widehat{FDR}$	&	$\widehat{TPR}$	\\
\midrule
{\em Overall selection}  	&		&		&		&		\\
elastic net 	&		&		&		&		\\
$\;\;$ $\alpha = 1$; min $\lambda$	&	--	&	35.38	&	0.74	&	1.00	\\
$\;\;$ $\alpha = 1$; min $\lambda$ + 1 s.e.	&	--	&	15.38	&	0.55	&	0.98	\\
$\;\;$ $\alpha = 0.5$; min $\lambda$	&	--	&	36.36	&	0.75	&	1.00	\\
$\;\;$ $\alpha = 0.5$; min $\lambda$ + 1 s.e.	&	--	&	17.32	&	0.58	&	0.99	\\
derandomized knockoff 	&	2	&	18.18	&	0.77	&	0.46	\\
sequential knockoff 	&	2	&	9.74	&	0.45	&	0.97	\\
\midrule
\multicolumn{4}{@{}l@{}}{\em Combination of selections at level 1  and level 2} 					&		\\
elastic net 	&		&		&		&		\\
$\;\;$ $\alpha = 1$; min $\lambda$	&	--	&	16.98	&	0.58	&	0.98	\\
$\;\;$ $\alpha = 1$; min $\lambda$ + 1 s.e.	&	--	&	6.27	&	0.34	&	0.94	\\
$\;\;$ $\alpha = 0.5$; min $\lambda$	&	--	&	19.24	&	0.61	&	0.98	\\
$\;\;$ $\alpha = 0.5$; min $\lambda$ + 1 s.e.	&	--	&	10.05	&	0.46	&	0.95	\\
derandomized knockoff 	&	(1, 1) 	&	3.24	&	0.22	&	0.91	\\
sequential knockoff 	&	(1, 1)	&	1.16	&	0.10	&	0.83	\\
\botrule
\end{tabular*}
\end{table}

As outlined in Section \ref{sec:methodproposal}, performing variable selection on the whole set of level 1 and level 2 variables simultaneously leads to very poor results, with the average number of false discoveries (FDR) ranging from $9.74$ (sparse sequential knockoff) to $36.36$ (elastic net with $\alpha = 0.5$ and $\lambda$ chosen to minimize the mean cross-validated error), and corresponding high false discovery proportions (FDR) from 0.45 (sparse sequential knockoffs) to 0.77 (derandomized knockoffs).
The benefits of the proposed splitting strategy for knockoff-based procedures is evident comparing the overall selection results for PFER=2 with the combined results of level 1 and level 2 selections with $PFER = 1$ at both levels (corresponding to a total nominal PFER equal to 2):  the MC mean of PFER is $1.16$ for the sparse sequential knockoff and $3.24$ for the derandomized knockoff, while the corresponding values obtained from the overall selection for $PFER = 2$ are $9.74$ and $18.18$, respectively.

The comparison between the two panels of Table \ref{tab:sim2} illustrates the advantage of our proposal, which involves carrying out the selection separately at both levels 1 and 2. The problem with overall selection is that the knockoffs of level 2 variables are not constant within clusters.

Further simulations, not reported here, show that the results of overall selection are even worse when raw level 1 variables are used (without centering and without adding cluster means). Indeed, in this setting, the estimated coefficients of the level 1 predictors are a combination of within- and between-effects. This is a further issue, in addition to the aforementioned problem of non-constant knockoffs within clusters for level 2 predictors.

\section{Conclusions}
\label{sec:final}

We proposed an extension of the knockoff filter for clustered data, specifically to address the selection of predictors that are constant within clusters (level 2 variables). Our methodological proposal involves splitting the variable selection problem into two independent tasks, specifically the selection of level 1 and level 2 predictors. This is achieved by centering level 1 predictors around their cluster means to make them uncorrelated with level 2 predictors.

We conducted a simulation study to evaluate the properties of our proposal, showing a clear benefit of the splitting approach in terms of PFER and FDR. Specifically, we compared two different implementations of the knockoff filter, finding that sparse sequential knockoffs outperform derandomized knockoffs. Even if the splitting strategy is not necessary for implementing the elastic net, it turns out to improve its performance, likely because it allows a penalty parameter specific to each hierarchical level.

A strength of the proposed approach is that it can be implemented by exploiting the available knockoff procedures, with the caveat of using weights in the selection of level 2 predictors for clusters of unequal sizes.
Further work is needed to handle binary level 1 predictors since this is not feasible within the available knockoff procedures.
Indeed, it is possible to center a binary variable around its cluster mean, which is the proportion of ones \citep{Yaremych:2023}. However, a centered binary variable takes two values within each cluster (no longer just zero and one), which are generally different across clusters. Thus, the available knockoff procedures require modification to address this issue. Note that there is no problem with binary predictors at level 2 since they do not have to be centered.

As the proposed knockoff filter works with clusters of unequal size,  it can accommodate missing responses within the clusters, such as dropout in repeated measures. However, the procedure requires the predictors to be fully observed. Otherwise, one can discard the units with missing values, which is generally not recommended \citep{rabe:2023}, or one can proceed with multiple imputation as suggested by \cite{bacci:2025}.
Further research is required to explore the impact on FDR and PFER of combining multiple imputation with the splitting approach for clustered data.

\bibliography{Knockoffs_multilevel_ArXiv_paper}

\end{document}